\documentclass[sigconf,nonacm]{acmart}

\usepackage{subfigure}
\begin{document}
\author{M V Sai Prakash}
\email{mukkamala.prakash@quantiphi.com}
\affiliation{%
  \institution{Applied Research, Quantiphi}
  \streetaddress{P.O. Box 1212}
  \city{Mumbai}
  \country{India}
  \postcode{43017-6221}
}

\author{Siddartha Reddy N}
\email{siddartha.reddy@quantiphi.com}

\affiliation{%
 \institution{Applied Research, Quantiphi}
  \streetaddress{P.O. Box 1212}
  \city{Bangalore}
  \country{India}
  \postcode{43017-6221}
}

\author{Ganesh Parab}
\email{ganesh.parab@quantiphi.com}
\affiliation{%
  \institution{Applied Research, Quantiphi}
  \streetaddress{P.O. Box 1212}
  \city{Mumbai}
  \country{India}
  \postcode{43017-6221}
}

\author{Varun V}
\email{varun.v@quantiphi.com}
\affiliation{%
  \institution{Applied Research, Quantiphi}
  \streetaddress{P.O. Box 1212}
  \city{Bangalore}
  \country{India}
  \postcode{43017-6221}
}

\author{Vishal Vaddina}
\email{vishal.vaddina@quantiphi.com}
\affiliation{%
    \institution{Applied Research, Quantiphi}
  \streetaddress{P.O. Box 1212}
  \city{Toronto}
  \country{Canada}
  \postcode{43017-6221}
}

\author{Saisubramaniam Gopalakrishnan}
\email{gopalakrishnan.saisubramaniam}
\email{@quantiphi.com}
\affiliation{%
 \institution{Applied Research, Quantiphi}
  \streetaddress{P.O. Box 1212}
  \city{Bangalore}
  \country{India}
  \postcode{43017-6221}
}
\authornote{Corresponding author}

\title{Synergistic Fusion of Graph and Transformer Features for Enhanced Molecular Property Prediction}


\renewcommand{\shortauthors}{Prakash, et al.} 


\begin{abstract}

Molecular property prediction is a critical task in computational drug discovery. While recent advances in Graph Neural Networks (GNNs) and Transformers have shown to be effective and promising, they face the following limitations: Transformer self-attention does not explicitly consider the underlying molecule structure while GNN feature representation alone is not sufficient to capture granular and hidden interactions and characteristics that distinguish similar molecules. To address these limitations, we propose SYN-FUSION, a novel approach that synergistically combines pre-trained features from GNNs and Transformers. This approach provides a comprehensive molecular representation, capturing both the global molecule structure and the individual atom characteristics. Experimental results on MoleculeNet benchmarks demonstrate superior performance, surpassing previous models in 5 out of 7 classification datasets and 4 out of 6 regression datasets. The performance of SYN-FUSION has been compared with other Graph-Transformer models that have been jointly trained using a combination of transformer and graph features, and it is found that our approach is on par with those models in terms of performance. Extensive analysis of the learned fusion model across aspects such as loss, latent space, and weight distribution further validates the effectiveness of SYN-FUSION. Finally, an ablation study unequivocally demonstrates that the synergy achieved by SYN-FUSION surpasses the performance of its individual model components and their ensemble, offering a substantial improvement in predicting molecular properties.\footnotetext{Accepted as poster at ACM-BCB 2023 and as workshop talk at CNB-MAC '23.}

\end{abstract}







\maketitle

\section{Introduction}

Molecular property prediction  \cite{MPP-dec-19} has rapidly evolved into an interdisciplinary field that leverages insights from chemistry, physics, and materials science. Predicting molecular properties is widely considered one of the most critical tasks in computational drug discovery. 
The ability to accurately predict molecular properties enables transformative applications in drug design, materials development, and reaction optimization \cite{ doi:10.1021/acscentsci.8b00507,OLIVEIRA2022863, https://doi.org/10.1002/inf2.12094}.
Early approaches in molecular property prediction include quantum mechanics-based mathematical models describing atomic and molecular behavior \cite{Woolley1998IsTA}, computational chemistry involving the study of chemical systems through computer simulations \cite{inproceedings}, and molecular mechanics and molecular dynamics \cite{Mouvet2022} involving the simulation of larger and more complex molecules.

With the advent of computational methods, fingerprint techniques such as Morgan Fingerprint \cite{Morgan1965}, Extended-Connectivity FingerPrints (ECFP) \cite{doi:10.1021/ci100050t} have been developed for efficient molecular representation. Statistical methods and machine learning models like Quantitative Structure-Property Relationships (QSPR) \cite{BLAKE1886} and Quantitative Structure-Activity Relationships (QSAR) \cite{article4} predict properties based on the molecular structure using these techniques. By utilizing large datasets to learn complex structure-property relationships \cite{LeCun2015-fu, Vamathevan2019-pl}, deep learning approaches surpass traditional methods  across various molecular tasks. Two such widely adopted approaches include modeling the molecule as (i) a sequence of atoms using either Simplified Molecular Input Line Entry System (SMILES) \cite{doi:10.1021/ci00057a005} or SELF-referencing Embedded Strings (SELFIES) \cite{Krenn_2022} and (ii) a graph-based structure using Graph Neural Networks (GNN) \cite{4700287}. 

Sequence-based molecular property prediction has witnessed significant growth in recent years \cite{10.1145/3307339.3342186,mgbert,DBLP:journals/corr/abs-2010-09885}. This approach leverages the inherent sequential nature of molecular structures and protein sequences to accurately predict properties, opening new possibilities in the realm of drug discovery and materials science. Graph neural networks (GNNs) model the natural representation of molecules as graphs and use neighborhood aggregation strategies to predict molecular properties such as solubility, toxicity, binding affinity, etc. \cite{pmlr-v70-gilmer17a,doi:10.1021/acs.jcim.9b00237} Related tasks, including molecular generation \cite{decao2022molgan}, reaction prediction \cite{mao2020molecular}, and molecular docking \cite{docking} are other applications of GNNs.

A pivotal challenge in property prediction is the limited availability of labeled data, an obstacle shared across different fields such as language and vision \cite{DBLP:conf/iccv/DoerschGE15,DBLP:conf/iclr/GidarisSK18}.
The success of self-supervised learning in image and text domains \cite{NEURIPS2020_f3ada80d,DBLP:conf/naacl/DevlinCLT19} has also been extended to molecular property prediction \cite{article22}. Contrastive learning \cite{pmlr-v119-chen20j} has been shown to be effective in learning better latent representations by pre-training a model to maximize the distance between positive and negative pairs from unlabeled data samples and learning downstream tasks with limited data \cite{NEURIPS2020_94aef384,DBLP:conf/iclr/HuLGZLPL20}. Masked Language Modelling (MLM) \cite{DBLP:conf/naacl/DevlinCLT19, lewis2020bart} has been adopted as a pre-training strategy in sequence-to-sequence and discriminative cheminformatics tasks \cite{Irwin_2022}. These approaches have proven beneficial for models to leverage the knowledge learned from larger datasets when fine-tuning smaller, focused tasks.

Both sequence-based transformers and graph-based models learn richer representations when they are initially pre-trained on large datasets of molecules in a self-supervised manner, followed by supervised fine-tuning on smaller datasets with specific properties of interest. This work investigates the benefits of fusing the pre-trained latent representations from both approaches and fine-tuning them towards the downstream task of property prediction.

The contributions of this work are as follows: (i) Learning the synergistic interaction between pre-trained features from graphs and transformers to create a more comprehensive molecular representation that captures both the global molecule structure and the characteristics of individual atoms, (ii) Conducting a detailed analysis of the learned synergistic fusion representation in various aspects, including loss, latent space, activation, and weights, through classification and regression case studies, and, (iii) an ablation study to showcase that the synergy effect of fusion is greater than the performances of the individual models and their ensemble.



\section{Related Works}
Molecular fingerprinting methods commonly used in cheminformatics and computational chemistry such as Extended-Connectivity FingerPrints (ECFP) \cite{doi:10.1021/ci100050t} encode the structural features of molecules into fingerprint representations for similarity-based analyses and machine learning tasks. ECFP generates a binary fingerprint for each molecule based on the structure, taking into account the connectivity of atoms and the presence of chemical groups. ECFP is fast but limited in its ability to capture the diversity of molecular structures, as it only considers the presence (1) or absence (0) of specific sub-structures within a molecule.
The Simplified Molecular Input Line Entry System (SMILES)  \cite{doi:10.1021/ci00057a005} is a linear representation designed to encode molecular structure in a machine-readable format. SMILES uses a distinctive set of characters to represent atoms, bonds, and functional groups within a molecule. Due to its machine-interpretability, SMILES has become a key molecular descriptor for training machine learning models. Advanced machine learning algorithms, including deep neural networks, can utilize SMILES strings to learn predictive models for various molecular properties. 
Transformer-based models \cite{10.1145/3307339.3342186,mgbert,DBLP:journals/corr/abs-2010-09885} have been employed for molecular property prediction, primarily relying on SMILES representations. Transformers utilize self-attention mechanisms that enable them to attend to different parts of the molecule and consider the relationships between atoms and bonds. However, while models such as Chemformer \cite{Irwin_2022} and X-MOL \cite{Xue2020.12.23.424259}, trained on SMILES data, have exhibited promising outcomes in classification and regression tasks, they have limitations in providing comprehensive insights into the underlying molecular structure, particularly in modeling the intricate connectivity patterns and spatial arrangements found in molecular graphs.

Graph Neural Networks (GNNs) have demonstrated significant potential in predicting molecular properties, as evidenced by previous studies \cite{pmlr-v70-gilmer17a,kipf2017semisupervised,DBLP:conf/iclr/XuHLJ19}. In GNNs, molecules are represented as graphs, with atoms serving as nodes and bonds as edges. These networks leverage the graph structure to learn representations of the molecules.
A widely adopted GNN-based approach is the Message Passing Neural Network (MPNN) \cite{pmlr-v70-gilmer17a}. MPNN utilizes a recursive message-passing mechanism to propagate information throughout the molecular graph structure. Another approach is the Graph Convolutional Network (GCN) \cite{DBLP:conf/iclr/KipfW17}, which uses graph convolution operations to learn node embeddings. 
GIN \cite{DBLP:conf/iclr/XuHLJ19} handles graph isomorphism (the similarity between two graphs despite differences in node labels or orderings) by employing an aggregation function that is independent of the ordering of nodes or edges. 

Molecules with similar overall structures can exhibit distinct functional groups or subtle variations that significantly impact their properties \cite{Pelzer2016}. Self-supervised learning has gained significant attention in the field of molecular property prediction, enabling models to learn meaningful representations from unlabeled molecular data.
Hu et al \cite{DBLP:conf/iclr/HuLGZLPL20} introduces innovative strategies for molecule graphs, encompassing pre-training at both the node and graph levels. 
Contrastive learning \cite{pmlr-v119-chen20j,oord2019representation} is a machine learning technique that learns data representations by comparing pairs of instances. Instances that are similar are pulled closer together, while instances that are dissimilar are pushed further apart. In molecular property prediction, contrastive learning-based methods involve encoding molecular structures as feature vectors and comparing these vectors using a contrastive loss function. An illustration of this is MolCLR \cite{Wang_2022}, which employs a contrastive loss function to learn representations of molecular structures that can subsequently be utilized for predicting molecular properties.
\textcolor{black}{MegaMolBART utilizes a transformer architecture based on BART \cite{lewis2020bart} and is trained for small molecule drug discovery. It comprises a bidirectional encoder and an autoregressive decoder. The pretraining of MegaMolBART is built upon the foundation of Chemformer \cite{Irwin_2022}} and uses Masked Language Modelling and augmentation of SMILES input. Recent efforts have focused on integrating graph information into Transformer models. GROVER \cite{DBLP:conf/nips/RongBXX0HH20} uses a graph multi-head attention network with node vectors obtained through a specialized Graph Neural Network (GNN). Graphormer \cite{ying2021do} is based on the standard Transformer, taking graph-structured encoded data directly as input and avoiding conversion to sequential formats. PharmHGT \cite{Jiang2023} utilizes various perspectives within the molecular graph for message passing, yielding distinct atom features, followed by attention aggregation for holistic molecule representation.




\begin{figure*}
    \centering
    \includegraphics[width=0.9\textwidth]{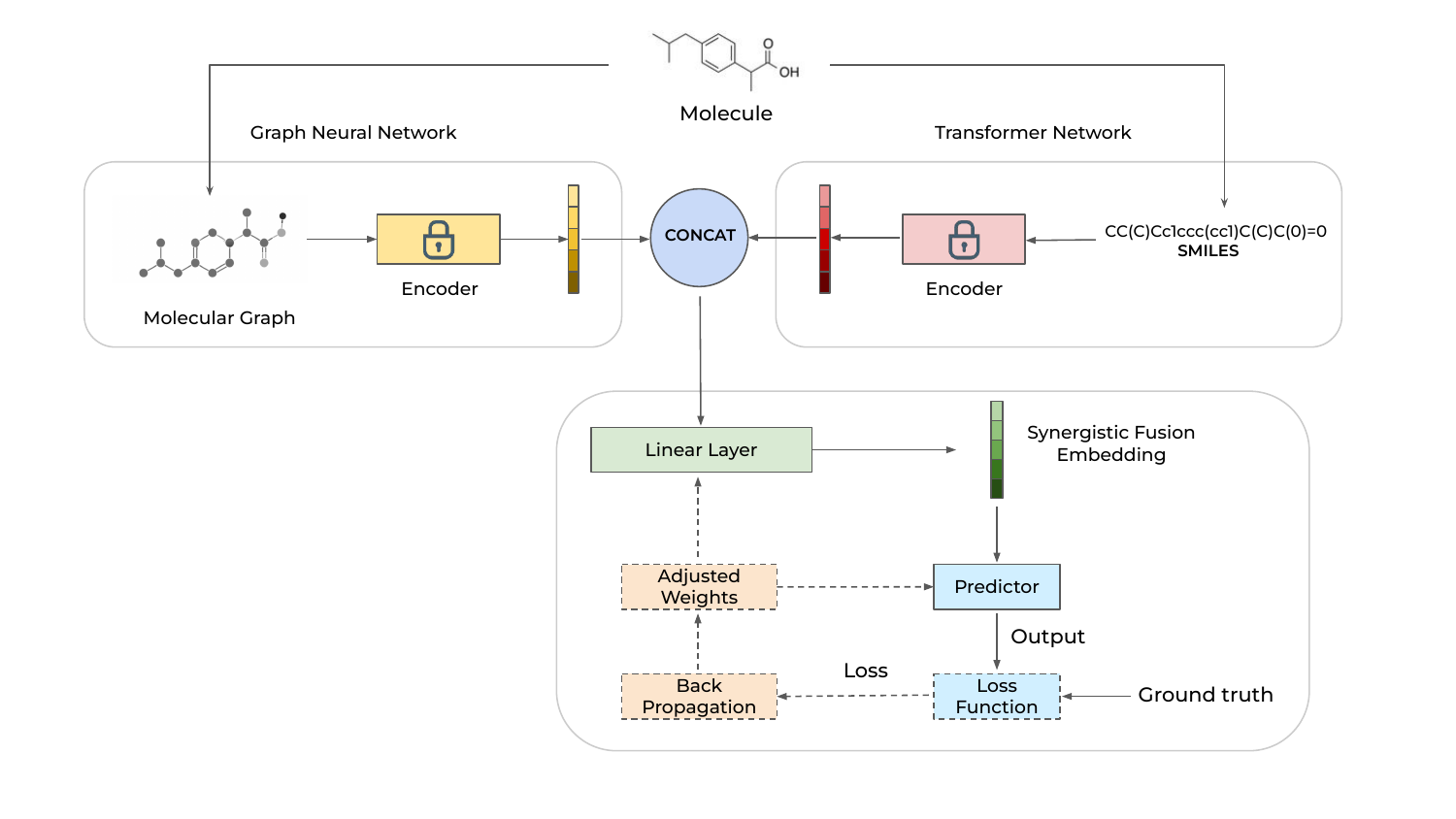}
    \caption{Framework of Synergistic Fusion: For a given molecule $M$, the molecular graph is processed through a pre-trained and frozen GNN model, generating a  feature vector $\mathbf{z}_{G}$. Simultaneously, the SMILES representation of $M$ is fed into a pre-trained and frozen transformer model, generating a feature vector  $\mathbf{z}_{T}$. The two vectors are concatenated and passed through a linear layer, yielding a fused representation $\textbf{z}_{F}$. This fused representation $\mathbf{z}_{F}$ is then utilized as input for a property predictor that estimates the molecule's properties $M$. The loss function computes the error between the predicted and the ground truth property values. The network parameters (weights) of both the predictor and the linear layer are adjusted/updated through back-propagation using the computed loss and gradients.}
    \label{fig:img1}
\end{figure*}


Transformers excel in learning complex relationships and hidden dependencies across the entire molecule, including interactions between atoms and bonds. However, they do not explicitly consider the underlying graph structure of the molecule. Alternatively, Graph Neural Networks (GNNs) offer a more appropriate framework for capturing the unique structural characteristics of molecules and have the potential to benefit by integrating transformer representations acquired through self-attention, effectively capturing long-range dependencies. Therefore, in this work, the pre-trained features from graphs and transformers have been combined, and the synergistic interaction between the two representations has been learned through fusion. The result is a more comprehensive molecular representation that effectively captures the global structure of molecules and the specific characteristics of individual atoms. Additionally, a detailed analysis of the learned synergistic fusion representation has been conducted through case studies involving classification and regression tasks. To the best of our knowledge, previous works have not explored using a combination of pre-trained features and the resulting synergistic interaction for molecular property prediction.


\section{Methodology}

We propose a novel approach to learning the synergistic interaction between pre-trained features from graphs and transformers, termed \textit{Synergistic Fusion} (SYN-FUSION). The overall framework is illustrated in Figure \ref{fig:img1}. The approach comprises two steps: In the first step, a single molecule is represented as a Graph and a SMILES string independently, and then encoded into two feature representations using a Graph Neural Network such as GIN and a Transformer based network such as MegaMolBART, respectively. In the second step, the two distinct features are concatenated and fused using a linear layer that learns the combined 'synergistic' information for enhanced downstream property prediction.

\subsection{Graph Isomorphism Network}
Graph Isomorphism Network (GIN) \cite{DBLP:conf/iclr/XuHLJ19} learns a permutation-invariant representation of each input graph, achieved by applying a series of message-passing operations to the nodes and edges of the graph.
Let G = (V, E) denote an undirected graph. 
Let $\mathbf{V}$ be the node feature matrix of the nodes in V and $\mathbf{E}$ be the edge feature matrix of the edges in E. The message-passing operation computes a new feature vector $\mathbf{h}$ for each node in V:

\begin{multline}
     \mathbf{h}_i^{(k+1)} = 
     \text{MLP}_{atom}^{(k+1)} \left( (1 + \epsilon^{(k)}) \, \mathbf{h}_i^{(k)} \right. \\
   \left. + \sum_{j \in \mathcal{N}(i)}  \Bigl(\mathbf{h}_j^{(k)} + \text{MLP}_{bond}^{(k+1) } (\mathbf{e}_{ij})\Bigl)\right) 
\end{multline}


where $\mathbf{h}_0 = V$, $\mathbf{e}_{ij}$ is the edge feature vector of edge $e$ $\in$ E connecting atoms $i$ and $j$, $k$ represents the $k_{th}$ layer, $MLP$ stands for multi layered perception, $MLP_{atom}^{(k+1)}$
 and $MLP_{bond}^{(k+1)}$
bond are the $(k + 1)$-th MLP layers on the atom- and
bond-level respectively, $\mathcal{N}(i)$ is the set of neighboring nodes of $i$, and $\epsilon^{(k)} $ is a learnable parameter that helps to avoid over-smoothing of the features.

The pooling operation aggregates the feature vectors of all nodes in the graph into a single vector:
\begin{equation}
\mathrm{pooling}\, (\mathbf{H}) = \mathrm{mean} \, (\mathbf{H})
\end{equation}
where $\mathbf{H} = [\mathbf{h}_1^{(kn)}, \mathbf{h}_2^{(kn)}, ..., \mathbf{h}_N^{(kn)}]$ is the matrix of node features at the final layer kn, N is the number of nodes in G and  $\mathrm{mean}(\cdot) $ is the element-wise mean operator.

The final output of the GIN network is obtained by passing the pooled feature vector through a linear layer:

\begin{equation}
\mathbf{z_{GIN}} = \mathrm{linear} \, (\mathrm{pooling}\,(\mathbf{H}))
\end{equation}
where $\mathbf{z_{GIN}}$ is the latent representation for the input graph G, and $\mathrm{linear}$ is a multi-layer perceptron.

By stacking multiple layers of message passing and pooling operations, GIN is able to learn a hierarchical representation of the input graph that is invariant to node ordering and is capable of capturing complex structural patterns.

\subsection{Transformer - MegaMolBART}

The Chemformer \cite{Irwin_2022} paper proposes a pre-training method for molecular property prediction based on the Bidirectional and Auto-Regressive Transformers (BART) \cite{lewis2020bart} architecture. 
MegaMolBART employs an identical configuration for pre-training i.e. Masked Language Modeling (MLM). The goal of MLM is to predict the masked tokens based on the context provided by the other tokens in the sequence. The pre-training process commences by transforming each molecule in the batch into a non-canonical SMILES representation that aligns with the specific molecule. The SMILES strings are subsequently subjected to random masking, tokenization, and embedding into a vector sequence. This modified sequence is then fed into the bidirectional encoder, while the autoregressive decoder is tasked with predicting the initial SMILES sequence based on the same right-shifted sequence. A fully-connected layer is employed to process the decoder's output, generating a distribution across the model's vocabulary. To obtain the latent feature from MegaMolBART, only the encoder component is required.
\begin{equation}
\textbf{z}_{\text{MMB}} = Encoder_{MMB}(x)
\end{equation}
where the bidirectional encoder of MegaMolBART, $Encoder_{MMB}$ takes a SMILES representation of the molecule $x$ as input and provides the corresponding latent representation $\textbf{z}_{\text{MMB}}$.



\subsection{Synergistic Fusion}
Synergy refers to the phenomenon where the combined effect of two or more substances is greater than the sum of their individual effects. It occurs when the interaction between the substances enhances or amplifies their overall impact \cite{10.3389/fphar.2017.00158}. 
 Let us consider two substances, denoted as $A$ and $B$, which each possess distinct effects represented by variables $X$ and $Y$, respectively. The combined effect resulting from the interaction of $A$ and $B$ can be represented as $Z$. If there exists synergy between substances $A$ and $B$, it can be expressed through the equation $ Z > X + Y$. This equation denotes that the combined effect $(Z)$ surpasses the summation of the individual effects $(X + Y)$, thereby indicating the presence of a synergistic interaction.




Let us define the latent embeddings from MegaMolBART encoder as $\mathbf{z}_{\text{MMB}}$ and the latent embeddings from GIN as $\mathbf{z}_{\text{GIN}}$.
\subsubsection{Classification}
For classification tasks using the cross-entropy loss, the objective function can be written as:
\begin{equation}
\mathcal{L}_{c} = -\frac{1}{N} \sum_{i=1}^{N} \sum_{j=1}^{C} y_{ij} \log(\hat{y}_{ij})
\end{equation}

where $N$ is the number of samples, $C$ is the number of classes, $y_{ij}$ is the true label for sample $i$ and class $j$, and $\hat{y}_{ij}$ is the predicted probability of sample $i$ belonging to class $j$.

We can now express the predicted probabilities as a function of the feature embeddings:

\begin{equation}
\hat{y}_{ij} = \text{softmax}\left(\mathbf{W}\left[\mathbf{z}_{\text{GIN}}^{(i)}, \mathbf{z}_{\text{MMB}}^{(i)}\right] + \mathbf{b}\right)_{ij}
\end{equation}

where $\mathbf{W}$ and $\mathbf{b}$ are the weight matrix and bias vector of the linear layer, $[\cdot,\cdot]$ denotes concatenation, and $\text{softmax}(\cdot)$ is the softmax function

Finally, we can write the  objective function for classification as:

\begin{equation}
\mathcal{L}_c = - \frac{1}{N} \sum_{i=1}^{N} \sum_{j=1}^{C} y_{ij} \log\left(\text{softmax}\left(\mathbf{W}\left[\mathbf{z}_{\text{GIN}}^{(i)}, \mathbf{z}_{\text{MMB}}^{(i)}\right] + \mathbf{b}\right)_{ij}\right)
\end{equation}
\subsubsection{Regression}
For regression tasks using the Mean Squared Error loss, the objective function can be written as: 
\begin{equation}
\mathcal{L}_r = \frac{1}{N} \sum_{i=1}^{N} (y_i - \hat{y}_i)^2
\end{equation}
where $N$ is the number of samples, $y_{i}$ is the true value for sample $i$, and $\hat{y}_{i}$ is the predicted value.

The predicted value $\hat{y}_i$ can be defined as:
\begin{equation}
\hat{y}_i = \Bigl(\mathbf{W}\left[\mathbf{z}_{\text{GIN}}^{(i)}, \mathbf{z}_{\text{MMB}}^{(i)}\right] + \mathbf{b}\Bigl)_{i}
\end{equation}

Finally, we can write the objective function for regression as:
\begin{equation}
\mathcal{L}_r = \frac{1}{N} \sum_{i=1}^{N} \biggl(y_i - \Bigl(\mathbf{W}\left[\mathbf{z}_{\text{GIN}}^{(i)}, \mathbf{z}_{\text{MMB}}^{(i)}\right] + \mathbf{b}\Bigl)_{i}\biggr)^2
\end{equation}

\section{Experiments}

\begin{table*}[!t]
    \caption{Classification Results on MoleculeNet datasets using scaffold split. SYN-FUSION approach outperforms the baselines in 5 out of 7 datasets. The best score for each dataset is indicated in bold and the second-best score is underlined.}
    \centering

    \begin{tabular}[width=0.8\linewidth]{llllllll}
    
    \hline
        & \multicolumn{6}{c}{\textbf{Classification (Higher is Better)}} \\ \hline
        \textbf{Metric} & \multicolumn{6}{c}{ROC-AUC (\%)}  \\ \hline
        \textbf{Dataset} & \textbf{BBBP} & \textbf{Tox21} & \textbf{ClinTox} & \textbf{HIV} & \textbf{BACE} & \textbf{SIDER} & \textbf{MUV} \\ 
        \textbf{Molecules} & 2,039 & 7,831 & 1,476 & 41,127 & 1,513 & 1,427 & 93,087 \\ 
        \textbf{Tasks} & 1 & 12 & 2 & 1 & 1 & 27 & 17 \\ \hline
        \textbf{RF} & 71.4 ± 0.0 & 76.9 ± 1.5 & 71.3 ± 5.6 & 78.1 ± 0.6 & \textbf{86.7 ± 0.8} & 68.4 ± 0.9 & 63.2 ± 2.3 \\
        \textbf{SVM} & 72.9 ± 0.0 & \textbf{81.8 ± 1.0} & 66.9 ± 9.2 & 79.2 ± 0.0 & \underline{86.2 ± 0.0} & 68.2 ± 1.3 & 67.3 ± 1.3 \\ 
        \textbf{GCN} \cite{DBLP:conf/iclr/KipfW17} & 71.8 ± 0.9 & 70.9 ± 2.6 & 62.5 ± 2.8 & 74.0 ± 3.0 & 71.6 ± 2.0 & 53.6 ± 3.2 & 71.6 ± 4.0 \\ 
        \textbf{GIN} \cite{DBLP:conf/iclr/XuHLJ19} & 65.8 ± 4.5 & 74.0 ± 0.8 & 58.0 ± 4.4 & 75.3 ± 1.9 & 70.1 ± 5.4 & 57.3 ± 1.6 & 71.8 ± 2.5 \\ 
        \textbf{D-MPNN} \cite{doi:10.1021/acs.jcim.9b00237} & 71.2 ± 3.8 & 68.9 ± 1.3 & 90.5 ± 5.3 & 75.0 ± 2.1 & 85.3 ± 5.3 & 63.2 ± 2.3 & 76.2 ± 2.8 \\ 
        \textbf{Hu et al.} \cite{DBLP:conf/iclr/HuLGZLPL20} & 70.8 ± 1.5 & 78.7 ± 0.4 & 78.9 ± 2.4 & \underline{80.2 ± 0.9} & 85.9 ± 0.8 & \underline{65.2 ± 0.9} & 81.4 ± 2.0 \\ 
 
        

         \textbf{MolCLR}\textsubscript{\textbf{GIN}}\cite{Wang_2022} & 73.9 ± 0.6  & 72.0 ± 0.7  & 88.6 ± 0.5 & 74.6 ± 1.6 & 77.9 ± 1.0 & 64.9 ± 0.5 & 83.8± 0.9  \\

         \hline

          \textbf{SYN-FUSION}\textsubscript{\textbf{Hu et. al}} & \underline{75.5 ± 0.7} & 73.8 ± 0.4 & \underline{94.6 ± 1.6}	& \textbf{83.7 ± 1.9} & 80.5 ± 1.1 & \textbf{69.9 ± 1.3} & \underline{88.9 ± 0.8} \\ 
           \textbf{SYN-FUSION}\textsubscript{\textbf{MolCLR}} & 74.2 ± 0.9 & 75.1 ± 0.6 & \textbf{94.7 ± 0.2}	& 76.3 ± 1.3 & 79.8 ± 0.4	& 65.0 ± 1.3 & \textbf{90.3 ± 1.3} \\
        \hline
    \end{tabular}
    \label{tab: Classification}
\end{table*}

\subsection{Data}
A series of experiments were conducted utilizing multiple molecular benchmarks obtained from MoleculeNet \cite{C7SC02664A}. These benchmarks encompass classification and regression tasks derived from diverse studies. To divide the datasets into training and testing sets, the scaffold split method \cite{Bemis1996-qz} was employed. A scaffold refers to a molecular substructure that exists within a group of molecules and serves to define a chemical series or class. To ensure the evaluation of the model's generalization ability, this procedure maintains chemical distinctiveness between the training and test sets. The training set comprised molecules possessing a specific scaffold, while the test set comprised molecules lacking that specific scaffold.
Using the scaffold split, the molecules in each dataset were divided into training, validation, and test sets, following an 8:1:1 ratio. The classification and regression results using scaffold split are provided in Table \ref{tab: Classification} and Table \ref{tab: Regression} respectively.
For a fair and consistent comparison between SYN-FUSION and Chemformer, X-Mol, and MolBERT models, random splitting was used instead of scaffold split during evaluation (Table \ref{Tab : Random Split}). This decision was due to the aforementioned models utilizing random splitting for their experiments, and adopting the same splitting methodology maintains methodological consistency across the comparative analysis.


\subsection{Configuration}



The SYN-FUSION framework utilized GIN and MegaMolBART as the GNN and Transformer architectures respectively. The experimental settings were adopted from \cite{Wang_2022}. For comparison purposes, two prior works that employed GIN were selected for fusion, namely, MolCLR \cite{Wang_2022} 
and Hu et. al \cite{DBLP:conf/iclr/HuLGZLPL20}. 
Adam was employed as the optimizer, maintaining a fixed learning rate of 0.001 across all models. A batch size of 32 was used for training the models on each dataset. The selection of an appropriate activation function is crucial as it significantly impacts the model's learning capacity. ReLU 
and Softplus 
activation functions were used as in \cite{Wang_2022}. For comparison purposes, the results using Softplus are presented in this work as it outperformed ReLU in terms of predictive performance. During the training phase, each model was learned for 100 epochs on the training set, with model checkpoints and early stopping based on validation loss. The model checkpoint having the lowest validation loss was used to evaluate the test set.

\subsection{Evaluation Metrics}

 ROC-AUC (\%) was used as the evaluation metric for all the classification datasets, following the recommendation by MoleculeNet. For regression datasets, Root Mean Square Error (RMSE) was employed as the metric for FreeSolv, ESOL, and Lipo datasets, while Mean Average Error (MAE) was used as the metric for the QM7, QM8, and QM9 datasets. For each method, the mean and standard deviation over three independent runs are presented.

\subsection{Baseline Methods}
SYN-FUSION underwent a comprehensive evaluation for molecular property prediction, comparing its performance with other GNN and Transformer baseline methods.
The evaluation process comprised of Random Forest (RF) and Support Vector Machine (SVM) which take molecular FPs as the input, GNN architectures that integrate edge features during the aggregation process such as GCN \cite{DBLP:conf/iclr/KipfW17} and GIN \cite{DBLP:conf/iclr/XuHLJ19}, capturing quantum interactions within molecules such as D-MPNN \cite{doi:10.1021/acs.jcim.9b00237}, node-level (self-supervised) and graph-level (supervised) pre-training approaches as described in Hu et. al \cite{DBLP:conf/iclr/HuLGZLPL20}, contrastive pre-training approach such as MolCLR \cite{Wang_2022}. Transformer-Graph Combination networks that use both self-attention and graph features for training such as Graphformer \cite{ying2021do}, pharmHGT \cite{Jiang2023} and GROVER \cite{DBLP:conf/nips/RongBXX0HH20}.
Due to inconsistencies between the reported results in the MolCLR paper and the reproduction using their provided code repository, the scores obtained from rerun using their code are presented and compared.

Transformer based models
such as Chemformer \cite{Irwin_2022}, X-Mol \cite{XMOL-2022}, and MolBERT \cite{MolBERT2021} were also included for comparison using random split as followed in their respective works. For a fair comparison, models that solely focus on 2D information were included, and models that incorporate 3D features like MGCN \cite{Lu_Liu_Wang_Huang_Lin_He_2019}, GEM \cite{Fang2022}, etc. were excluded.

\subsection{Results}
\begin{table*}[!t]
    \caption{Regression Results on MoleculeNet datasets using scaffold split. SYN-FUSION approach outperforms baselines in 4 out of 6 datasets. The best score for each dataset is indicated in bold and the second-best score is underlined.}
    \centering
    \begin{tabular}[width=0.8\linewidth]{llll|lll}
        
        \hline
        \multicolumn{7}{c}{\textbf{Regression (Lower is Better)}} \\ \hline
        \textbf{Metric} & \multicolumn{3}{c}{RMSE} & \multicolumn{3}{c}{MAE} \\ \hline
        \textbf{Dataset} & \textbf{FreeSolv} & \textbf{ESOL} & \textbf{Lipo} & \textbf{QM7} & \textbf{QM8} & \textbf{QM9} \\ 
        \textbf{Molecules} & 642 & 1,128 & 4,200 & 6,830 & 21,786 & 130,829 \\ 
        \textbf{Tasks} & 1 & 1 & 1 & 1 & 12 & 8 \\  \hline

        \textbf{RF} & 2.10 ± 0.22 & 1.07 ± 0.19 & 0.88 ± 0.04 & 122.7 ± 4.2 & 0.0423 ± 0.0021 & 16.061 ± 0.019 \\

        \textbf{SVM} & 3.14 ± 0.00 & 1.50 ± 0.00 & 0.82 ± 0.00 & 156.9 ± 0.0 & 0.0543 ± 0.0010 & 24.613 ± 0.144 \\

        \textbf{GCN} \cite{DBLP:conf/iclr/KipfW17} & 2.87 ± 0.14 & 1.43 ± 0.05 & 0.85 ± 0.08 & 122.9 ± 2.2 & 0.0366 ± 0.0011 & 5.796 ± 1.969 \\
        \textbf{GIN} 
        \cite{DBLP:conf/iclr/XuHLJ19} & 2.76 ± 0.18 & 1.45 ± 0.02 & 0.85 ± 0.07 & 124.8 ± 0.7 & 0.0371 ± 0.0009 & 4.741 ± 0.912 \\ 
        \textbf{D-MPNN} \cite{doi:10.1021/acs.jcim.9b00237} & 2.18 ± 0.91 & 0.98 ± 0.26 & \textbf{0.65 ± 0.05} & 105.8 ± 13.2 & \textbf{0.0143 ± 0.0022} & 3.241 ± 0.119 \\ 
        \textbf{Hu et al.} \cite{DBLP:conf/iclr/HuLGZLPL20} & 2.83 ± 0.12 & 1.22 ± 0.02 & 0.74 ± 0.00 & 110.2 ± 6.4 & 0.0191 ± 0.0003 & 4.349 ± 0.061 \\ 
        
       \textbf{MolCLR}\textsubscript{\textbf{GIN}} \cite{Wang_2022} &2.81 ± 0.03 ~ &1.29 ± 0.01 & 0.79 ± 0.00~ & 92.3 ± 1.5 ~ & 0.0187 ± 0.0012 & 2.933 ± 0.053 ~ \\

        \hline

         \textbf{SYN-FUSION}\textsubscript{\textbf{Hu et. al}} & 3.13 ± 0.02 &  0.96 ± 0.007 & \underline{0.70 ± 0.01} & \underline{67.5 ± 1.3} & 0.0187 ± 0.0041 & \underline{1.947 ± 0.096}   \\ 

         \textbf{SYN-FUSION}\textsubscript{\textbf{MolCLR}} & \textbf{2.08 ± 0.04 }& \textbf{0.89 ± 0.02} & 0.72 ± 0.01 & \textbf{64.8 ± 1.4} & \underline{0.0181 ± 0.0001} & \textbf{1.892 ± 0.042}   \\ 
         \hline
    \end{tabular}
    \label{tab: Regression}
\end{table*}

\begin{table*}[!t]
    \caption{Classification and Regression Results using random split. SYN-FUSION approach outperforms baselines on all datasets. The best score for each dataset is indicated in bold and the second-best score is underlined.}
    \centering

    \begin{tabular}[width=0.8\linewidth]{lllll|lll}
    
    \hline
        ~ & \multicolumn{4}{c} {\textbf{Classification }} & \multicolumn{3}{c}{\textbf{Regression}}\\ \hline
        \textbf{Metric} & \multicolumn{4}{c}{ROC-AUC (\%)} & \multicolumn{3}{c}{RMSE} \\ \hline
        \textbf{Dataset} & \textbf{BBBP} & \textbf{BACE} & \textbf{ClinTox} & \textbf{HIV} & \textbf{ESOL} & \textbf{Lipo} & \textbf{FreeSolv} \\ 
        \textbf{Molecules} & 2,039 & 1,513 & 1,476 & 41,127 & 1,128 & 4200 & 642 \\ 
        \textbf{Tasks} & 1 & 1 & 2 & 1 & 1 & 1 & 1 \\ \hline
        \textbf{Chemformer} \cite{Irwin_2022} & - & - & - & - & 0.633 & 0.598 & 1.230  \\
        \textbf{X-Mol} \cite{XMOL-2022} & \underline{96.0} & 87.2 & \underline{99.3} & \underline{79.8} &  0.578 & 0.596 & 1.108 \\
        \textbf{MolBERT} \cite{MolBERT2021} & 87.5 & - & 92.3 & - & 0.531 & \underline{0.561} & 0.948\\
        \hline
        \textbf{SYN-FUSION}\textsubscript{\textbf{MolCLR}} & \textbf{96.5 ± 0.3} & \textbf{90.2 ± 0.4} & \textbf{99.5 ± 0.1} & \textbf{84.2 ± 0.8} & \textbf{0.496 ± 0.06} & \textbf{0.534 ± 0.02} & \textbf{0.876 ± 0.04} \\
        
         \textbf{SYN-FUSION}\textsubscript{\textbf{Hu et. al}} & 95.5 ± 0.2 & \underline{88.4 ± 0.5} & 98.3 ± 1.2 & \underline{81.2 ± 0.4} & \underline{0.529 ± 0.21} & 0.729 ± 0.03 & \underline{0.937 ± 0.12} \\
       \hline
    \end{tabular}
    \label{Tab : Random Split}
\end{table*}

\begin{table*}
\caption{Comparison with Transformer-Graph combination networks on classification and regression datasets from MoleculeNet. The best score for each dataset is indicated in bold and the second-best score is underlined.}
 \begin{tabular}[width=0.8\linewidth]{llll|lll}
    
    \hline
        ~ & \multicolumn{3}{c} {\textbf{Classification }} & \multicolumn{3}{c}{\textbf{Regression}}\\ \hline
        \textbf{Metric} & \multicolumn{3}{c}{ROC-AUC (\%)} & \multicolumn{3}{c}{RMSE} \\ \hline
         \textbf{Dataset} &  \textbf{ClinTox} & \textbf{HIV}  & \textbf{SIDER} & \textbf{FreeSolv} & \textbf{ESOL} & \textbf{Lipo} \\ \hline

         \textbf{GROVER} \cite{DBLP:conf/nips/RongBXX0HH20} &   94.4 ± 2.1 & 68.2 ± 1.1 & 65.8 ± 2.3& \underline{1.99 ± 0.07} &1.10 ± 0.18 & 0.82 ± 0.01  \\

         \textbf{Graphormer} \cite{ying2021do} & 88.1 ± 3.8 & 78.9 ± 0.9 & 62.0 ± 1.2 & 2.09 ± 0.75  ~ & 0.93 ± 0.04 &  1.10 ± 0.39 \\
            
         \textbf{PharmHGT} \cite{Jiang2023} &   94.5 ± 0.4  &  \underline{80.6 ± 0.2}  &   \underline{66.9 ± 1.6} &  \textbf{1.70 ± 0.52} & \textbf{0.84 ± 0.05} &  \textbf{0.64 ± 0.04} \\ \hline

           \textbf{SYN-FUSION}\textsubscript{\textbf{Hu et. al}} & \underline{94.6 ± 1.6}	& \textbf{83.7 ± 1.9} & \textbf{69.9 ± 1.3}  & 3.13 ± 0.02 &  0.96 ± 0.007 & \underline{0.70 ± 0.01} \\ 
           \textbf{SYN-FUSION}\textsubscript{\textbf{MolCLR}}  & \textbf{94.7 ± 0.2}	& 76.3 ± 1.3 &  65.0 ± 1.3 & 2.08 ± 0.04 & \underline{0.89 ± 0.02} & 0.72 ± 0.01 \\

            \end{tabular}
    \label{Tab : Graphormer}
\end{table*}

\subsubsection{Classification}
The performance of SYN-FUSION was assessed on seven classification datasets. The comparison with GNN baselines using scaffold split is provided in Table \ref{tab: Classification}. SYN-FUSION exhibited superior performance in 5 out of 7 such as  ClinTox, HIV, SIDER, and MUV.
SYN-FUSION has a relative improvement of (6.63, 0.4)\% on BBBP, (-6.2, 4.3)\% on Tox21, (19.89, 6.88)\% on ClinTox, (4.36, 2.27)\% on HIV, (-6.2, 2.4)\% on BACE, (7.2, 0.15)\% on SIDER, (9.21, 7.75)\% on MUV datasets when comparing against its non-fusion counterparts Hu et. al \cite{DBLP:conf/iclr/HuLGZLPL20} and MolCLR \cite{Wang_2022} approaches respectively. Also on ClinTox and SIDER datasets, SYN-FUSION outperforms the state-of-the-art model by 4.64\% and 2.2\% respectively.
The results using random split are provided in Table \ref{Tab : Random Split}. SYN-FUSION demonstrated an improvement over X-Mol \cite{XMOL-2022} by 0.8\%, 3.4\%, 0.3\%, and 6\% on BBBP, BACE, ClinTox, and HIV datasets respectively. The comparison involving Transformer-Graph combination networks is presented in Table \ref{Tab : Graphormer}. The SYN-FUSION model demonstrated performance improvements of 2\%, 3.8\%, and 4.4\% on ClinTox, HIV, and SIDER datasets, respectively, compared to the previous best.
\subsubsection{Regression} 

The performance of SYN-FUSION was assessed on six regression benchmarks and the corresponding results are presented in Table \ref{tab: Regression}. SYN-FUSION  surpassed the performance of baseline methods on 4 out of the 6 datasets namely FreeSolv, ESOL, QM7, and QM9. 
These datasets encompass a diverse range of molecular properties, thereby providing a comprehensive evaluation of the fusion approach's capabilities.
SYN-FUSION has a relative improvement of (-10.6, 25.97)\% on FreeSolv, (21.3, 31)\% on ESOL, (5.4, 8.86)\% on Lipo, (38.74, 29.79)\% on QM7,  (2.09, 3.2)\% on QM8, (55.23, 35.49)\% on QM9 datasets when comparing on Hu et. al \cite{DBLP:conf/iclr/HuLGZLPL20} and MolCLR \cite{Wang_2022} approaches respectively. 
Notably, the SYN-FUSION model demonstrated significant improvements over the previous best, achieving a remarkable 4.3\% improvement on the ESOL dataset. 
In the random split experiments (Table \ref{Tab : Random Split}) SYN-FUSION showcased substantial enhancements on ESOL, Lipo, and FreeSolv datasets, achieving improvements of 6.59\%, 4.81\%, and 7.59\% respectively, surpassing the performance of all the baselines. The comparison involving Transformer-Graph combination networks is presented in Table \ref{Tab : Graphormer}. Although SYN-FUSION's performance does not surpass that of methods specifically trained using both graphs and transformers, our approach is comparable. Moreover, our method demonstrates greater practicality due to its efficiency in terms of reduced training time and computational resources required.

\section{Analysis on Synergistic Fusion}
We conducted an extensive analysis of our synergistic fusion approach, SYN-FUSION, to evaluate its performance from both quantitative and qualitative perspectives. The analysis covered various aspects, such as examining its latent space, activation maps, loss interpolation, and weight distribution. All experiments conducted in this section compared SYN-FUSION as the synergistic model with MolCLR in GNN and MegaMolBART in Transformer as its individual models.
\begin{figure*}
  \centering

  \subfigure[t-SNE SYN-FUSION]{
    \includegraphics[width=0.3\linewidth]{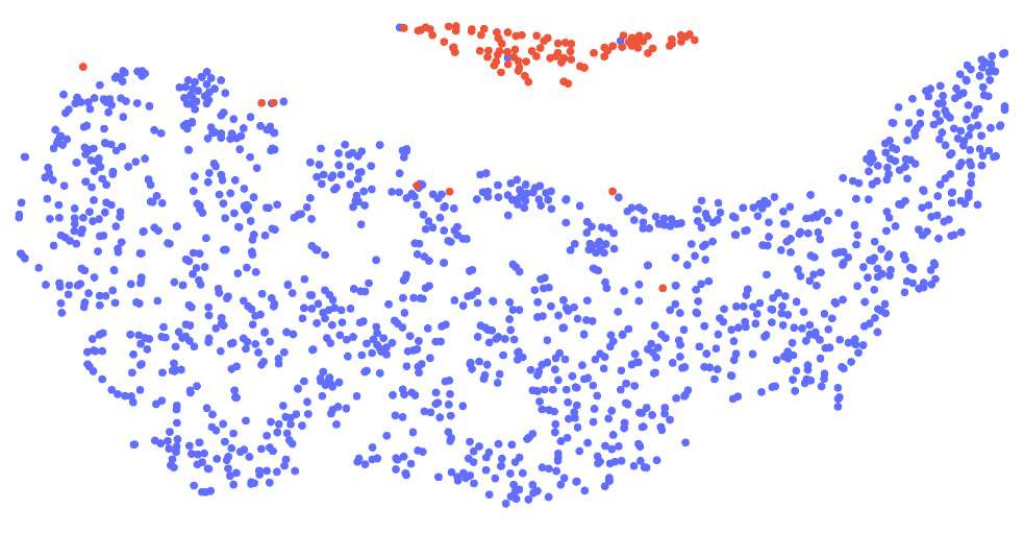}
    \label{fig:subfigureA}
  }
  \hfill
  \subfigure[t-SNE MolCLR]{
    \includegraphics[width=0.3\linewidth]{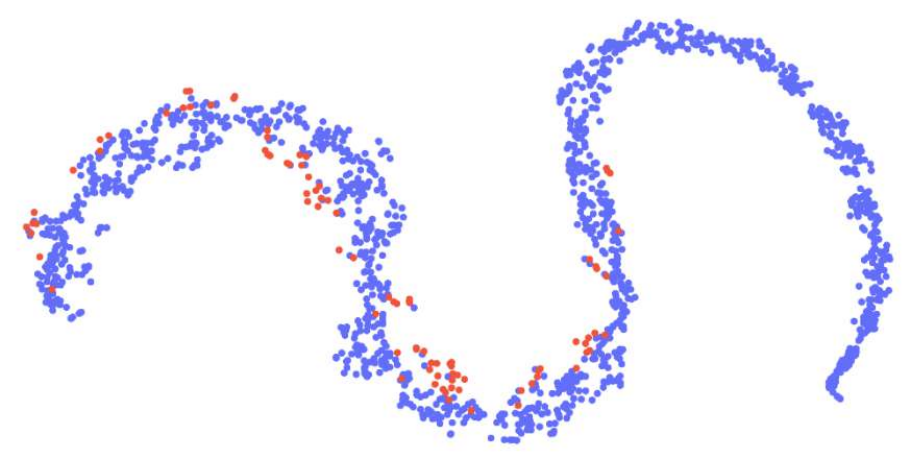}
    \label{fig:subfigureB}
  }
  \hfill
  \subfigure[t-SNE MegaMolBART]{
    \includegraphics[width=0.3\linewidth]{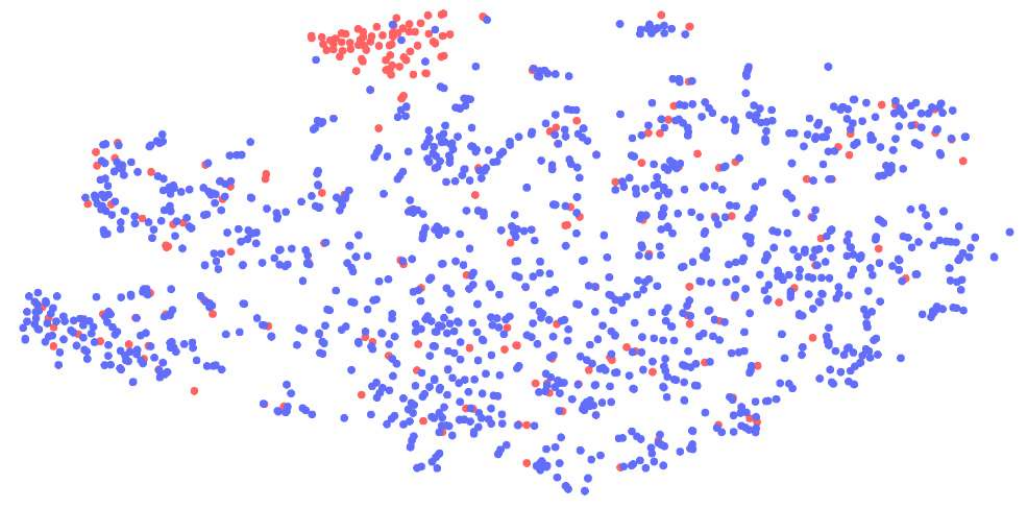}
    \label{fig:subfigureC}
  }

  \caption{ Latent Space Visualization for SYN-FUSION, MolCLR, and MegaMolBART using t-SNE on ClinTox dataset. The red color represents toxic molecules, while the blue color represents non-toxic molecules. Compared to the other two methods, SYN-FUSION demonstrates a pronounced ability to achieve a clear separation between toxic and non-toxic molecules.}
\label{t-SNE}
\end{figure*}

\begin{figure*}
  \centering

  \subfigure[Confusion Matrix - SYN-FUSION]{
    \includegraphics[width=0.3\linewidth]{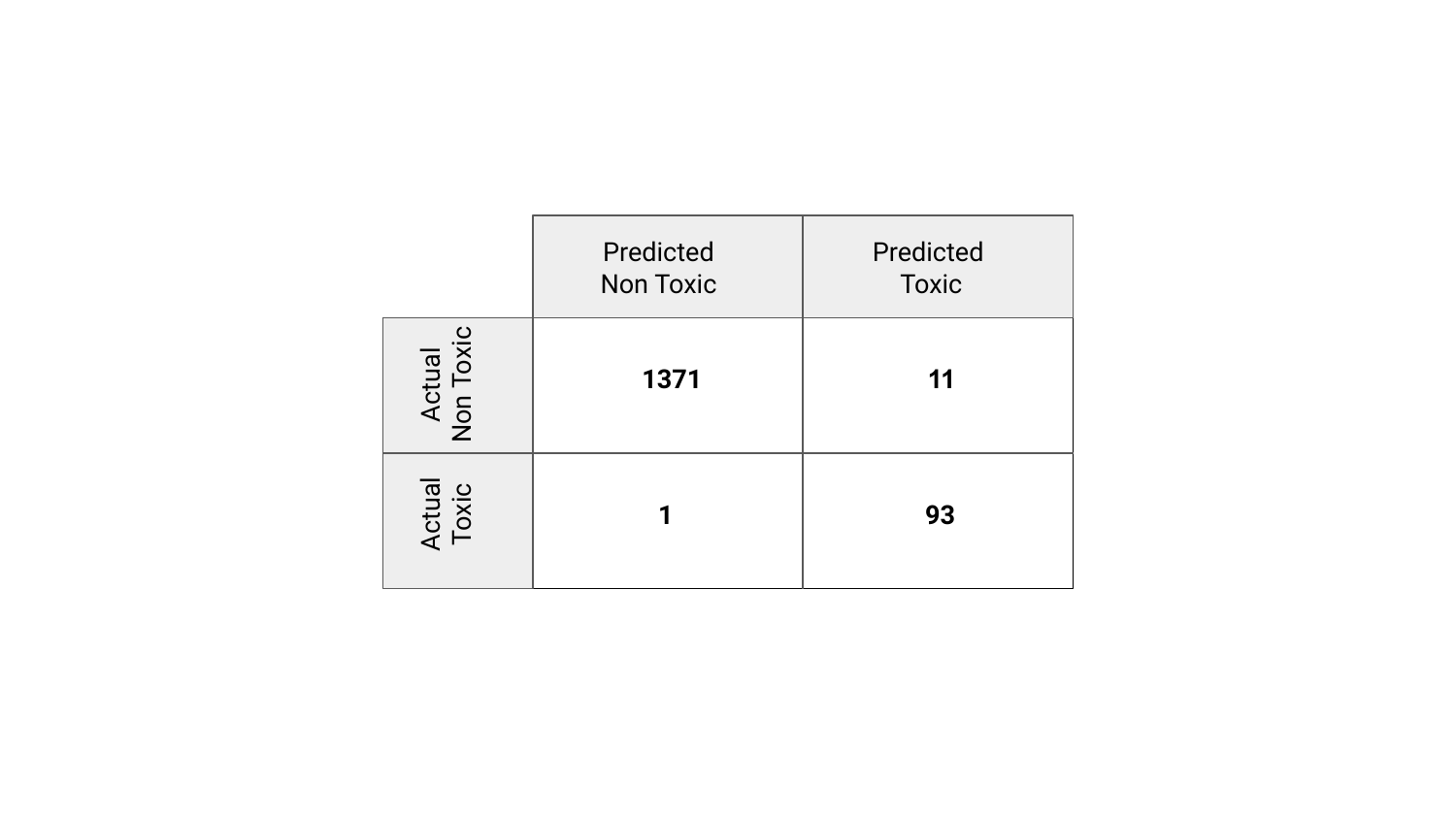}
    \label{fig: Fusion CM}
  }
  \hfill
  \subfigure[Confusion Matrix - MolCLR]{
    \includegraphics[width=0.3\linewidth]{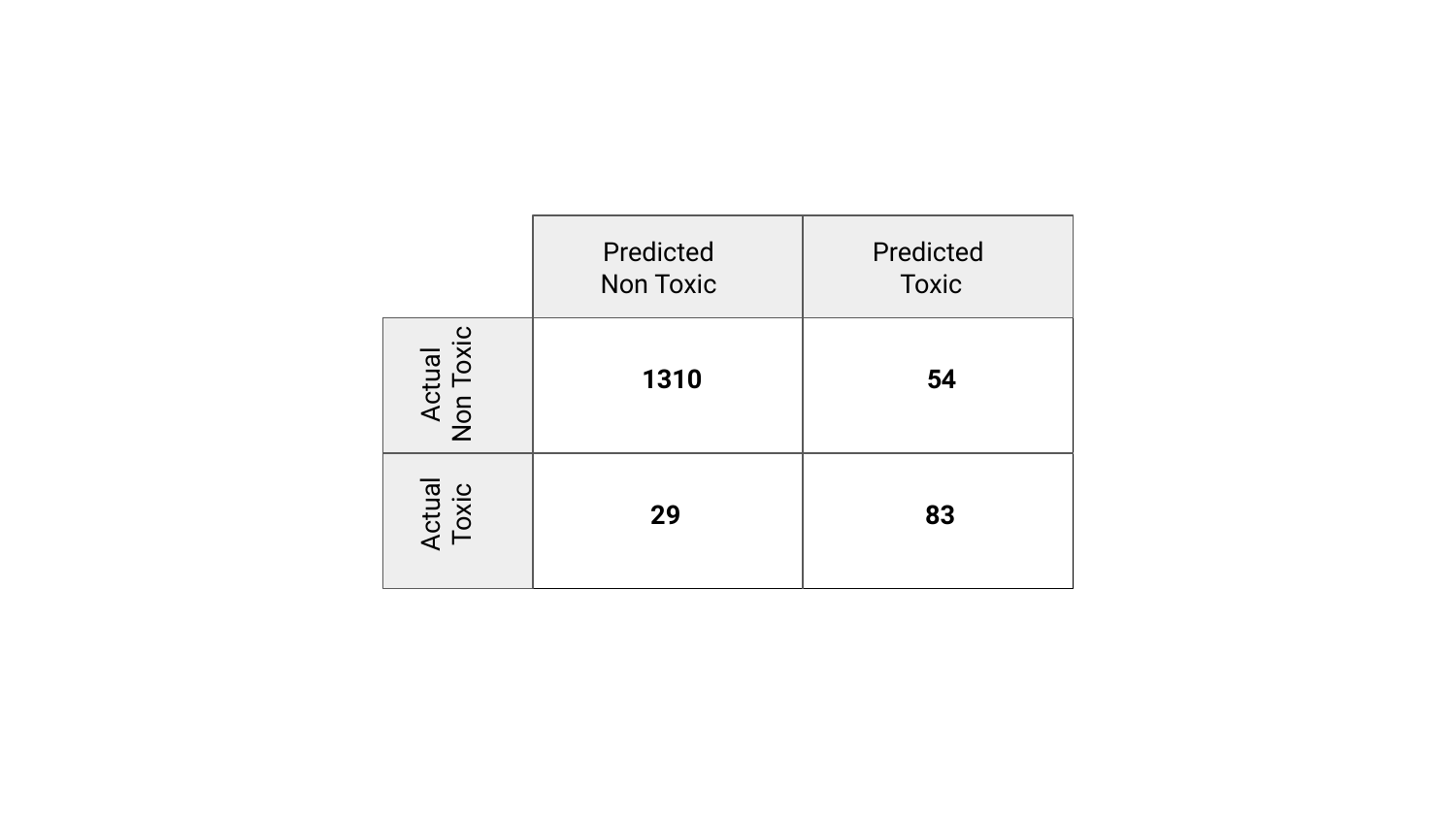}
    \label{fig: MOLCLR CM} 
  }
  \hfill
  \subfigure[Confusion Matrix - MegaMolBART]{
    \includegraphics[width=0.3\linewidth]{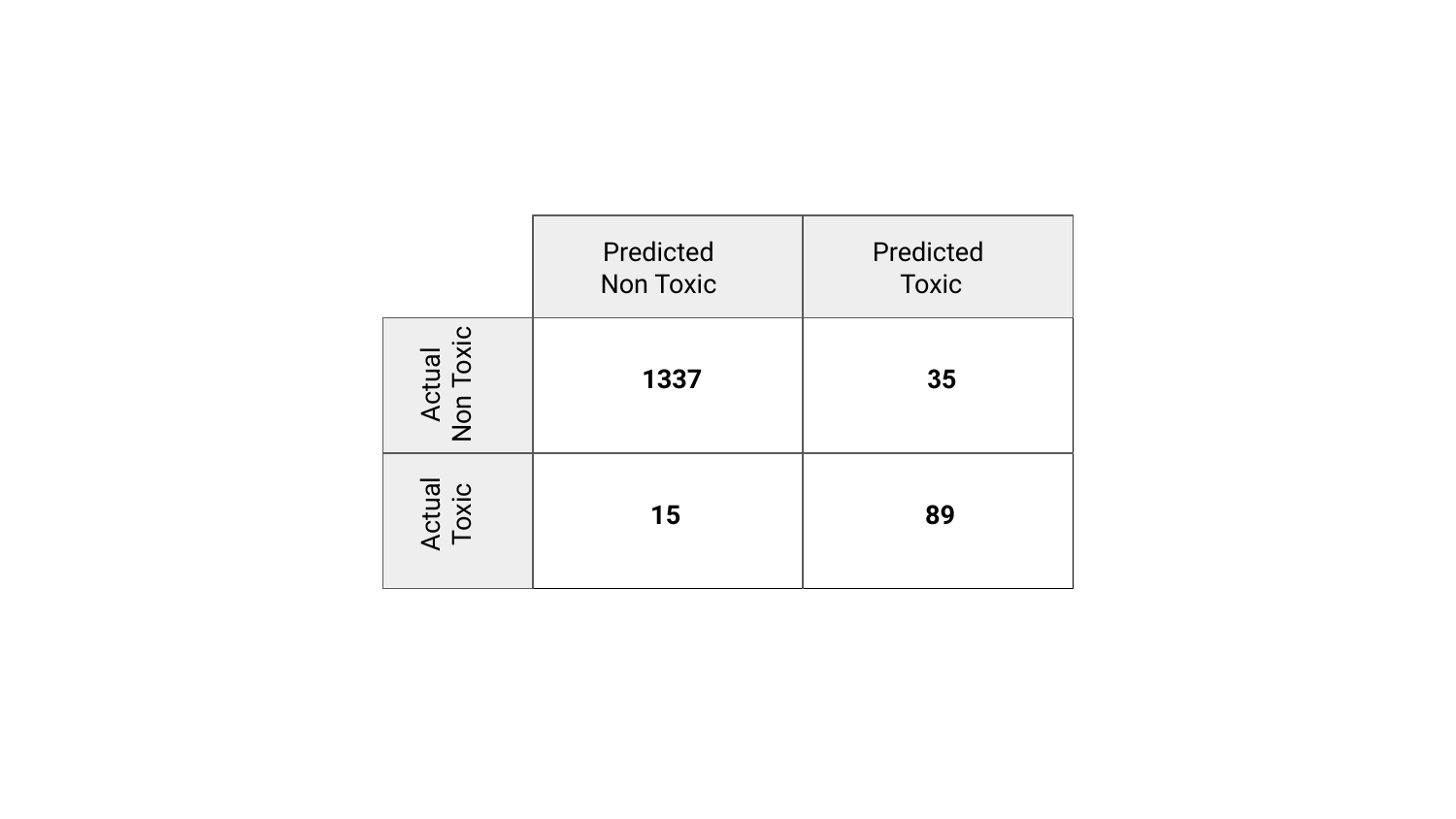}
    \label{fig: MMB CM}
  }

  \caption{Confusion Matrix of SYN-FUSION, MolCLR, and MegaMolBART on ClinTox dataset. SYN-FUSION demonstrates enhanced efficiency in molecular classification and achieves lower rates of both false negatives and false positives.}
\label{fig: sCM}
\end{figure*}

\begin{figure*}
  \centering

  \subfigure[Activation Map - SYN-FUSION]{
    \includegraphics[width=0.3\linewidth]{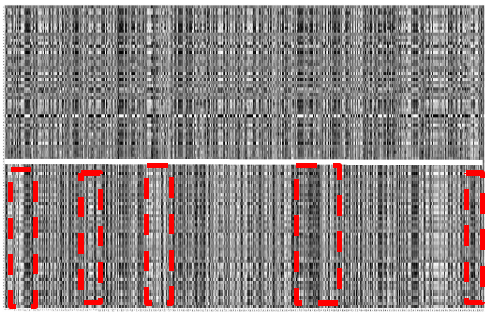}
    \label{fig:Activation Fusion}
  }
  \hfill
  \subfigure[Activation Map - MolCLR]{
    \includegraphics[width=0.3\linewidth]{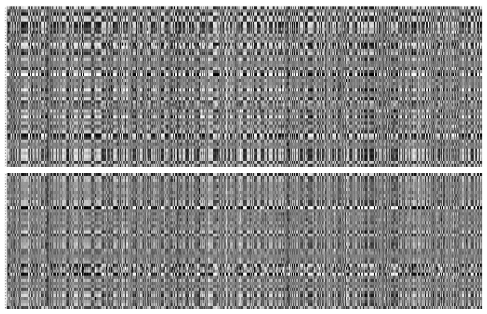}
    \label{fig:Activation MOLCLR}
  }
  \hfill
  \subfigure[Activation Map - MegaMolBART]{
    \includegraphics[width=0.3\linewidth]{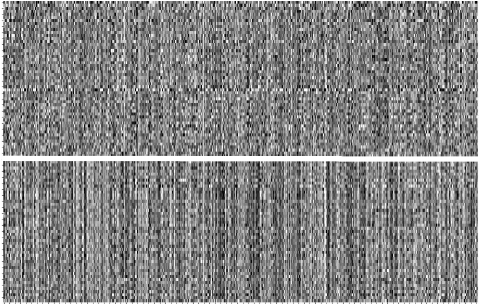}
    \label{fig:Activation MMB}
  }

  \caption{Activation Maps of SYN-FUSION, MolCLR, and MegaMolBART on ClinTox dataset.  The activation map is split into two parts, and each row of a part comprises a 1D vector of the pre-final layer. The resulting barcode per part is obtained by stacking 50 such samples belonging to the same class. The top part represents the activations of 50 samples from the toxic class while the bottom part contains 50 from the non-toxic class. The activation map of SYN-FUSION unveils notable patterns for both toxic and non-toxic molecules which are highlighted in the color red.}
\label{Activation}
\end{figure*}

\subsection{Latent Space Visualization}


Latent space visualization offers valuable insights into the encoded representations learned by a model, enabling a better understanding of the learned distribution and patterns. t-SNE \cite{JMLR:v9:vandermaaten08a} plots were generated to compare the latent space representations of the proposed SYN-FUSION model and its individual components (MolCLR and MegaMolBART) on the ClinTox classification dataset, providing qualitative visualization for comparison as shown in Figure \ref{t-SNE}, with subfigures \ref{fig:subfigureA}, \ref{fig:subfigureB}, and \ref{fig:subfigureC} displaying the t-SNE plots of the embeddings derived from SYN-FUSION, MolCLR, and MegaMolBART, respectively. The red and blue points represent the projection of toxic and non-toxic molecule samples respectively. Figure \ref{fig:subfigureA} has a clear separation between the two classes, as the toxic samples cluster together at the top while the non-toxic molecules appear at the bottom. This observation indicates the successful learning and encoding of discriminative features pertaining to toxic and non-toxic molecules by SYN-FUSION, and the model's ability to make accurate predictions regarding the toxicity of new molecules.
In contrast, Figure \ref{fig:subfigureB} suggests that the latent representations of toxic and non-toxic molecules in MolCLR are intermingled instead of having a separation.
This finding implies that MolCLR may face difficulties in accurately classifying molecules as toxic or non-toxic. MegaMolBART (Figure \ref{fig:subfigureC}) exhibits improved discrimination between toxic and non-toxic molecules, although there are still scattered instances of toxic molecules among the non-toxic ones, and the level of separation is not as pronounced as in SYN-FUSION. Confusion matrices obtained on evaluation of 1476 molecule samples from ClinTox presented in Figures \ref{fig: Fusion CM}, \ref{fig: MOLCLR CM}, and \ref{fig: MMB CM} indicate that better separation leads to fewer false predictions, and SYN-FUSION made fewer incorrect predictions in distinguishing between toxic and non-toxic molecules when compared to MolCLR and MegaMolBART.

\subsection{1-D Activation Maps}
Activation maps help to identify similar patterns across samples belonging to the same class and enable the model to distinguish between classes, leading to effective decision-making. To observe any learned patterns between toxic and non-toxic molecules, 100 samples from ClinTox were considered in equal proportions (50 toxic / 50 non-toxic), and 1-D activation maps were generated using the last layer of SYN-FUSION, MolCLR, and MegaMolBART models. 
Figure \ref{fig:Activation Fusion} showcases the stacked 1-D activation maps of SYN-FUSION, revealing distinct and clear activation patterns across samples in both classes, indicating that the model has learned to focus on relevant features for effective classification.
In contrast, the activation maps of MolCLR in Figure \ref{fig:Activation MOLCLR} lack well-defined patterns, and it is difficult to differentiate the toxic from the non-toxic class. Activation maps generated by MegaMolBART, as depicted in Figure \ref{fig:Activation MMB}, demonstrate intermediate characteristics between the two models, revealing a few discernible patterns that are slightly more noticeable compared to MolCLR but not as prominent as SYN-FUSION.

\subsection{Loss Interpolation}
\label{loss-interpolation}
\begin{figure}
    \begin{subfigure}
    \centering
\includegraphics[width=0.9\linewidth]{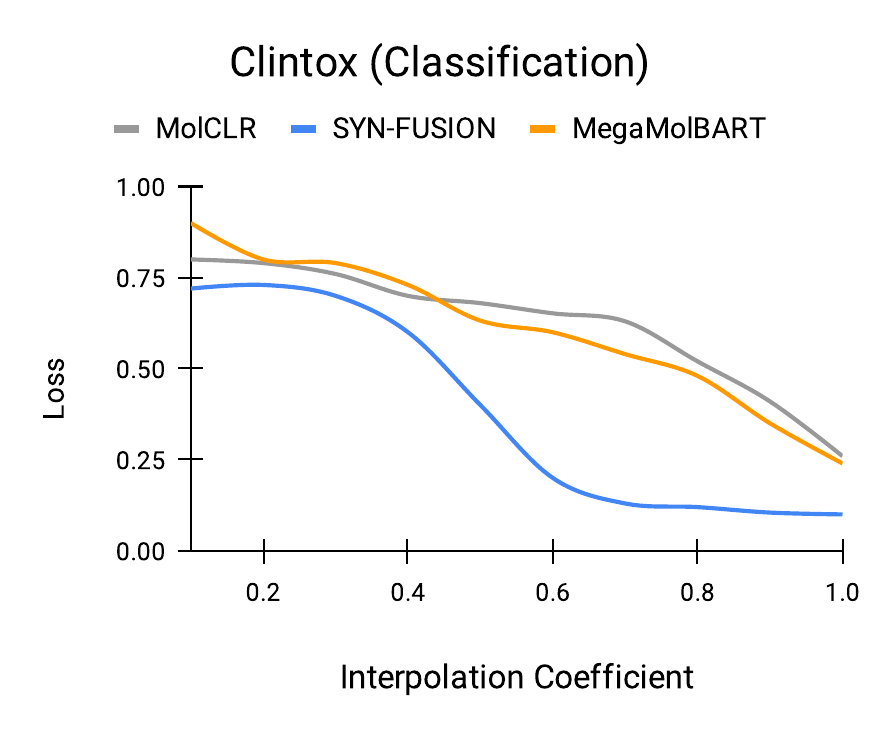}
    \end{subfigure}
    \begin{subfigure}
    \centering
\includegraphics[width=0.9\linewidth]{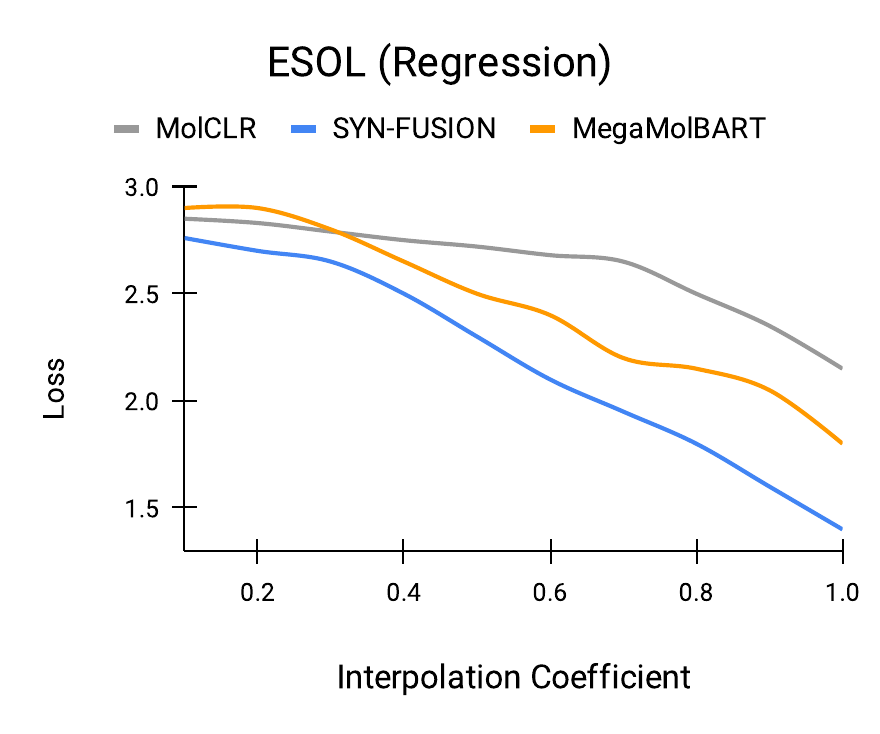}
    \end{subfigure}
    \caption{Loss Interpolation plots of SYN-FUSION, MolCLR, and MegaMolBART on ClinTox and ESOL Datasets. The loss trajectory of the SYN-FUSION displayed a significantly high level of monotonicity.}
    \label{fig:loss interpolation}
\end{figure}

Loss interpolation plots offer a concise visualization of the transition between different loss values, providing valuable insights into the behavior and convergence of optimization over the course of training. Notably, the presence of Monotonic Linear Interpolation in a model's loss trajectory signifies that the optimization of tasks is relatively easier \cite{DBLP:journals/corr/GoodfellowV14}. 
Notable differences are observed in the loss trajectories of the models under comparison, as shown in Figure \ref{fig:loss interpolation}. Specifically, the loss curve of SYN-FUSION displays a remarkably high level of monotonicity, suggesting a smoother and more consistent optimization process, in contrast to the  loss curves of MolCLR and MegaMolBART. Moreover, SYN-FUSION has lower initial and final loss values compared to the other two models, providing additional evidence of easier and better optimization.
These findings substantiate the effectiveness of synergistic fusion surpassing individual models in terms of optimization and convergence.

\subsection{Weight Histograms}
\begin{figure}[!t]
  \centering
\includegraphics[width=0.9\linewidth]{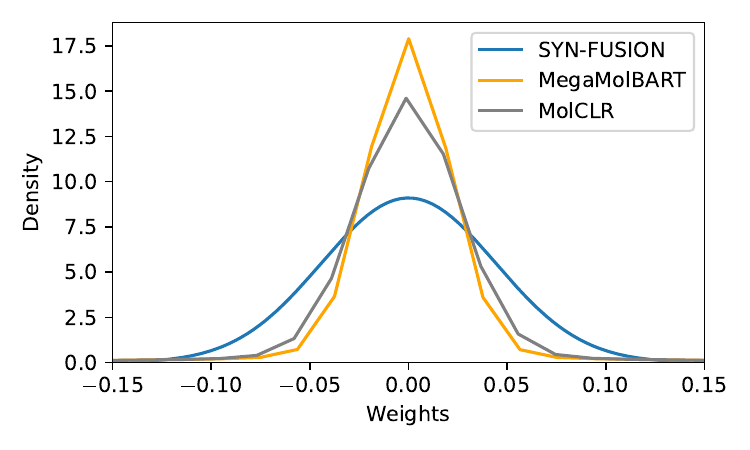}
  \caption{Weight Histograms of SYN-FUSION, MolCLR, and MegaMolBART on ClinTox dataset. In SYN-FUSION, the distribution of weights extends across a wide range of magnitudes, encompassing both high and low values.}
  \label{fig: weight hist}
  
\end{figure}
A weight histogram shows the distribution of weights within a model, providing insights into the range and frequency of different weight values. Small weights (values close to 0.0) tend to yield sharper minimizers and exhibit greater sensitivity to perturbations \cite{NEURIPS2018_a41b3bb3}. 
Conversely, weight distribution with uniform variance (both positive and negative values) leads to flatter minima and contributes to better generalization.
In light of this finding, the weight distributions of the final layer of SYN-FUSION, MolCLR, and MegaMolBART were investigated after completion of training, and the histogram of weights is presented in Figure \ref{fig: weight hist}. SYN-FUSION produces higher magnitude (both range and density) weights compared to MolCLR and MegaMolBART, indicating that fusion improves generalization and helps in easier and faster optimization. The impact of this phenomenon can be observed in the loss interpolation discussed in Section \ref{loss-interpolation}  Figure \ref{fig:loss interpolation} where SYN-FUSION demonstrates a favorable initialization denoted by a lower initial loss value, undergoes a rapid minimization of the loss ending with a significantly lower final loss value.

\section{Ablation Study}
\begin{figure}[!t]
    \begin{subfigure}
    \centering
\includegraphics[width=0.8\linewidth]{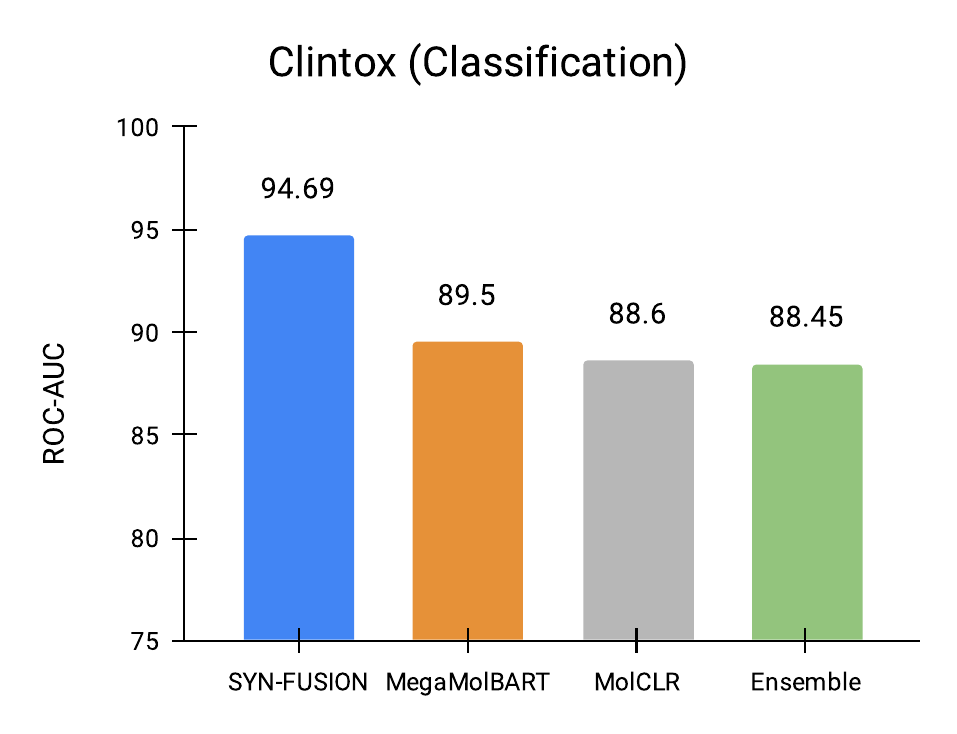}
    \end{subfigure}
    \begin{subfigure}
    \centering
\includegraphics[width=0.8\linewidth]{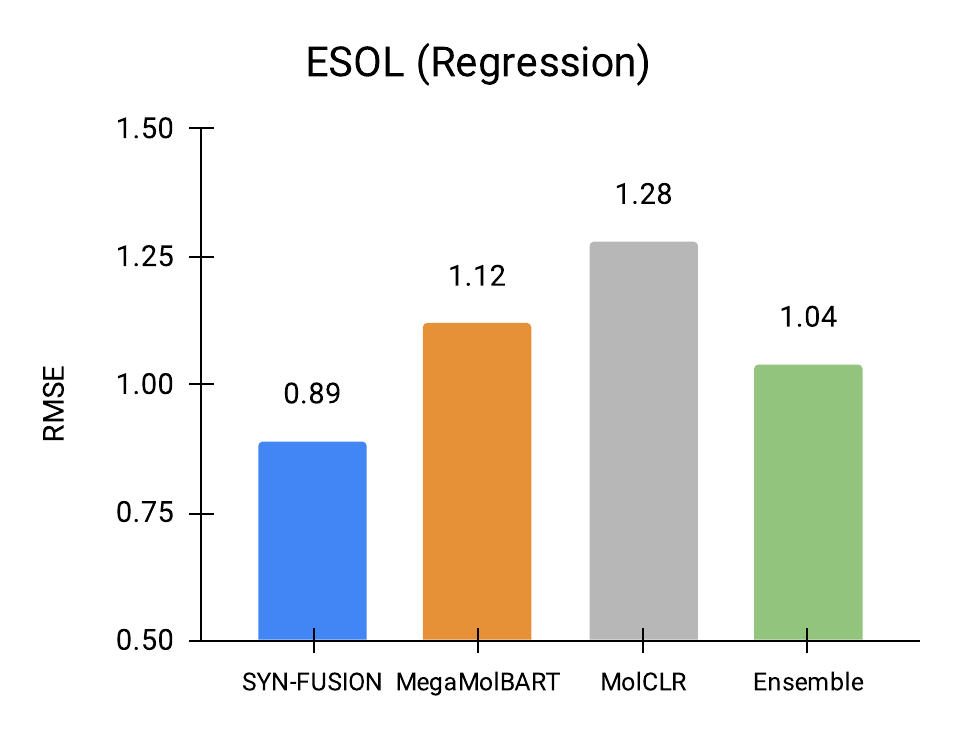}
    \end{subfigure}
    \vspace{-0.2cm}
    \caption{Performance of SYN-FUSION, MolCLR, and MegaMolBART on ClinTox and ESOL Datasets. A higher value of ROC-AUC indicates better performance, while a lower value of RMSE suggests better performance.}
    \label{Ablation}
\end{figure}
In order to experimentally verify the impact of synergy, we conducted a comparative analysis between the combined effect (represented by SYN-FUSION) and the individual models (MolCLR and MegaMolBART), as well as their (sum) ensemble, on both classification and regression tasks. In the ensemble approach, we handled the predictions generated by each individual model differently depending on the task at hand. For classification, if both models provided identical predictions, we retained the prediction as is. However, in cases where the models offered differing predictions, we considered the prediction with higher confidence. For regression, we computed the average of the two individual model predictions. The results are illustrated in Figure \ref{Ablation}. In the absence of SYN-FUSION, the AUC\% drops from 94.69\% by 5.19\%-6.24\% on ClinTox, and the RMSE increases from 0.89 by 0.15-0.39 on ESOL. This demonstrates the synergy effect - the combined effect achieved through fusion is greater than the individual models and their ensemble.

\vspace{-0.2cm}
\section{Discussion and Conclusion}



We present SYN-FUSION, a novel approach that synergistically combines pre-trained features from Graph Neural Networks (GNNs) and Transformers to create a comprehensive molecular representation. Our method effectively captures both the global structure of molecules and the characteristics of individual atoms, addressing the limitations of existing approaches.
Experimental results on various molecular benchmarks demonstrate the superior performance of SYN-FUSION compared to previous models in both classification and regression tasks. Furthermore, a detailed analysis of the learned fusion model provides insights into its effectiveness through aspects such as loss, activation, and weight distribution. The conducted ablation study demonstrates that the fusion approach outperforms individual models and their ensemble, offering a substantial improvement in predicting molecular properties.
This work contributes to the advancement of molecular representation techniques, providing a promising solution for accurate molecule property prediction and generalization within the vast chemical space. We believe that the presented findings will make a substantial contribution to the field, and hold great potential for applications in drug discovery and chemical research.

\bibliographystyle{ACM-Reference-Format}
\bibliography{references}

\end{document}